\documentclass[doublecol,linenumbers]{epl2} 

\title{Anatomy of nuclear matter fundamentals}
\shorttitle{Anatomy of nuclear matter fundamentals} 

\author{S. K. Patra \and S. K. Biswal \and S. K. Singh \and M. Bhuyan}
\shortauthor{S. K. Patra \etal}

\institute{Institute of Physics, Sachivalaya Marg, Bhubaneswar-751 005, 
India.}

\abstract{
The bridge between finite and infinite nuclear system is analyzed for 
the fundamental quantities like binding energy, density, compressibility, 
giant monopole excitation energy and effective mass of both nuclear 
matter and finite nuclei systems. It is shown quantitatively that by 
knowing one of the fundamental property of one system one can estimate 
the same in its counter part, only approximately.}

\begin{document}

\maketitle

The empirical values of nuclear matter binding energy, compressibility 
and  density are essential quantities for the estimation of nuclear 
observables. These are the center of attraction from the inception of 
nuclear physics. Recently, it is shown that the density of finite nucleus 
can be derived from the nuclear matter estimation \cite{mario09}. In the 
present letter, we want to show that not only the density of finite 
nucleus is connected with its infinite counter part, but also the 
compressibility which is an indispensable ingredient for nuclear study 
is also connected with each other. It is already a settled issue that 
the neutron skin thickness $S=R_n-R_p$ has a direct correlation with 
the equation of state (EOS) and asymmetric energy coefficient, hence 
consequently with the neutron star. Not only the neutron skin, recently 
a large number of observables, such as the stiffness parameter $Q$ and 
the $L$ coefficient are interconnected to one another in a very 
correlated way. These are not the only correlation between the 
fundamental nuclear properties with the properties of equation of states,
recently it is shown that there exhibit a correlation with the degree 
of neutron-proton asymmetry in nuclei with the experimental data on 
pygmy dipole strength \cite{palit07}. These correlations are anatomically 
shown by the advent of the relativistic mean field (RMF) formalism 
originally suggested by Johnson and Teller \cite{teller55} and later 
on developed by many others \cite{miller72,walecka74,boguta77,serot86}.

The advantage of the relativistic mean field formalism 
\cite{serot97,tang96} is its easy application to both finite and infinite
nuclear matter and the term-by-term analysis of the Lagrangian connected 
to the physical observables of the nuclear system. Taking into account 
a few parameters and the masses of mesons and nucleons, one can reproduce 
the finite nuclei properties through out the periodic chart \cite{patra0}. 
The inclusion of relativistic frame-work and the meson-nucleon interaction, 
the model predicts phenomena in a much fundamental levels. In the present 
paper, we would like to use the well known Lagrangian of Boguta and Bodmer
\cite{boguta77}, along with the cross coupling addition, as suggested by
Todd-Rutel et al.  \cite{todd05},  with a few parameter sets, which are 
well tested in most of the regions of the nuclear landscape. In this 
Lagrangian, the nonlinear couplings of the $\sigma-$meson is included,  
which generates analogous effect of the three body interaction due to 
it's off-shell meson couplings, which is essential for the saturation 
properties \cite{fujita57,steven01,schiff50}. These terms give the long 
range repulsion of nuclear force generated by the singlet-singlet and 
triplet-triplet $nn-$interaction. Therefore the two non-linear terms 
($\frac{1}{3}b\phi^3$ and $\frac{1}{4}c\phi^4$) are not only mere 
addition to the Lagrangian, rather it is essential to add in the 
Lagrangian to get a proper description of nuclear system. The other 
cross coupling $R^2V^2$ is important for EOS and neutron-rich matter. 
The nucleon-meson interacting Lagrangian for a many-body nucleonic
system is \cite{patra0,boguta77,tang96,todd05}:
\begin{eqnarray}
{\cal H}&= &\sum_i \varphi_i^{\dagger}
\bigg[ - i \vec{\alpha} \cdot \vec{\nabla} +
\beta m^* + g_{v} V + \frac{1}{2} g_{\rho} R \tau_3 \nonumber\\
&+&\frac{1}{2} e {\cal A} (1+\tau_3) \bigg] \varphi_i 
+ \frac{1}{2} \left[ (\vec{\nabla}\phi)^2 + m_{s}^2 \phi^2 \right]
+\frac{1}{3} b \phi^3\nonumber\\
&+& \frac{1}{4} c \phi^4
-\frac{1}{2} \left[ (\vec{\nabla} V)^2 + m_{v}^2 V^2 \right] +\Lambda_v(R^2V^2)^2\nonumber\\
&-& \frac{1}{2} \left[ (\vec{\nabla} R)^2 + m_\rho^2 R^2 \right]
- \frac{1}{2} \left(\vec{\nabla}  {\cal A}\right)^2 . 
\end{eqnarray}
Here $m$, $m_s$, $m_v$ and $m_{\rho}$ are the masses for the nucleon
(with $m^*=m-g_s\phi$ being the effective mass of the nucleon), ${\sigma}-$,
$\omega-$ and ${\rho}-$mesons, respectively and ${\varphi}$ is the Dirac 
spinor. The field for the ${\sigma}$-meson is denoted by ${\phi}$, for 
${\omega}$-meson by $V$, for ${\rho}$-meson by $R$ ($\tau_3$ as the 
$3^{rd}$ component of the isospin)  and for photon by $A$. $g_s$, $g_v$, 
$g_{\rho}$ and $e^2/4{\pi}$=1/137 are the coupling constants for the 
${\sigma}$, ${\omega}$, ${\rho}$-mesons and photon respectively. The 
quantities such as $b$ and $c$ are the non-linear coupling constants 
for ${\sigma}$ mesons, and $\Lambda_v$ is the crossed coupling constant 
for $\rho-$ and $\omega-$ mesons.

The above Lagrangian is used to determine the fundamental quantities 
of nuclear matter (BE/A, $J$, K) at different density $\rho$ both in 
(i) quantal and (ii) semi-classical approximations. The mean field 
(Hartree) approach of meson field is assumed in the quantal case and 
in semi-classical approximation, the scalar density ($\rho_s$) and 
energy density ($\cal{E}$) are calculated using relativistic Thomas-Fermi
(RTF)  and relativistic extended Thomas-Fermi (RETF) formalisms. The 
RETF is the ${\hbar}^2$ correction to the RTF, where the gradient of 
density is taken care. This term of the density takes care of the 
variation of the density and involves more in the surface properties.

We calculate the nuclear matter compressibility $K$, effective mass 
$m^*$ and binding energy per particle as a function of density using 
the RMF models. The similar quantities are also evaluated for some 
specific finite nuclei in the frame-work of same Hartree approximation.
Since the collective properties of nuclei, such as giant monopole, 
quadrupole and dipole resonances do not depend much on the internal 
structure of nuclei, we use the RTF and RETF techniques to calculate 
the values, whenever required. The recently developed scaling and 
constrained calculations will be used to evaluate the giant monopole 
resonances\cite{patra01} and all other quantities will be estimated 
by the relativistic Hartree approximation 
\cite{boguta87,patra91,lala97,todd05}.

\bigskip
\begin{table*}
\begin{center}
\caption{\label{tab:table1}{The binding energy per nucleon ($^{eos}$BE/A),
compressibility modulus $^{eos}K$, asymmetry coefficient $^{eos}J$ obtained
from nuclear matter equation of state (EOS) compared with the values of 
finite nuclei. The results of nuclear matter are listed at the density 
of finite nuclei using FSUGold parameter set. $^SK_A$ and $^CK_A$ are the 
compressibility of finite nucleus obtained from scaling and constrained 
calculations in MeV, respectively.
}}
\bigskip
\begin{tabular}{|c|c|c|c|c|c|c|c|c|c|c|c|}
\hline
Nucleus&$\rho_A$&$^{S}K_{A}$&$^{C}K_{A}$&$^{eos}K_{\rho}$&$^{eos}J$&BE/A&$^{eos}$BE/A\\
\hline
$^{40}$P& 0.0780& 123.400&100.36 &102.53&21.88&8.933&12.997\\
$^{40}$S& 0.0780& 127.028&112.15 &102.53&21.88&8.375&12.997\\
$^{40}$Ca& 0.0780& 130.93&123.15 &102.53&21.88&10.589&12.997\\
$^{112}$Sn& 0.0933& 147.23&140.26 &129.86&24.54&11.854&14.358\\
$^{116}$Sn& 0.0920& 147.11&139.71 &127.64&24.33&11.723&14.263\\
$^{120}$Sn& 0.0944& 146.62&138.66 &132.00&24.75&11.575&14.452\\
$^{124}$Sn& 0.0950& 145.83&137.14 &134.35&24.95&11.397&14.452\\
$^{208}$Pb& 0.0990& 147.37&134.57 &138.42&25.37&11.919&14.721\\
\hline
\end{tabular}
\end{center}
\end{table*}
\begin{table*}
\begin{center}
\caption{\label{tab:table2}{The calculated sum rule weight-age
${{\sqrt{{m_1/m_{-1}}}}}$ compared with the recently measured data 
using FSUGold parameter set. The results are also compared with the 
theoretical calculations of pairing plus MEM.}}
\begin{tabular}{|c|c|c|c|c|c|c|c|c|c|c|c|}
\hline
Nucleus &\multicolumn{3}{c|}{${\sqrt{{m_1/m_{-1}}}}$}
&\multicolumn{2}{c|}{$\Gamma$}\\ \hline
&pairing+MEM&RETF&Expt.& RETF & Expt.\\ \hline
$^{204}Pb$&13.4&13.6&13.7$\pm$0.1&2.02&3.3$\pm$0.2 \\
$^{206}Pb$&13.4&13.51&13.6$\pm$0.1&2.03& 2.8$\pm$0.2 \\
$^{208}Pb$&13.4&13.44&13.5$\pm$0.1&2.03&3.3$\pm$0.2  \\
\hline
\end{tabular}
\end{center}
\end{table*}
The mean density of a given nucleus is written by the fitting formula
\cite{mario09} $\rho_A=\rho_0-\frac{\rho_0}{(1+cA^{1/3})}$. The 
corresponding $K_A$, $J_A$ and BE/A for the finite nucleus are noted 
down in the Hartree or RETF formalisms. Again all these quantities are 
calculated from the equation of state and compared in Table I. The 
nuclear matter compressibility $K_{\infty}$ is not directly measured 
experimentally, actually, the energy $E_M$ of the GMR of finite nuclei 
is measured. It is convenient to write this energy in terms of the 
compressibility $K_A$ for a finite nucleus of mass number A as
\begin{eqnarray}
E_M=\sqrt{\frac{\hbar^2K_A}{M<r^2>}},
\end{eqnarray}
where $<r^2>$ is the rms matter radius and $M$ the mass of the nucleon.
The finite nucleus compressibility $K_A$ usually parametrized by means of
a leptodermous expansion that is similar to the liquid drop model mass
formula \cite{blaizot80}:
\begin{eqnarray}
K_A&=&K_{\infty} + K_{sf}^{-1/3}+K_{vs}I^2 \nonumber \\
&&+K_{Coul}Z^2A^{-4/3}+....
\end{eqnarray}
where $I=(N-Z)/(N+Z)$ is the neutron excess. Thus, the compressibility 
of a finite nucleus is an admixture of its volume, surface, asymmetric 
and Coulomb parts. In the semiclassical calculations, all these four 
contributions estimated combinedly. However, we can separate these 
individual terms taking the help of TF and ETF formalisms with some 
additional working conditions. For example, a nucleus with N=Z and 
switching off the Coulomb contribution, only contributions come from 
volume and surface. The surface part can be estimated by taking the 
difference of compressibility between the TF and ETF formalisms, as 
ETF gives the surface contribution on top of the TF formalism. The 
finite nucleus compressibility $K_A$ compared with the one obtained 
from the infinite nuclear matter in Table I.  From the Table, it is 
clear that the compressibility calculated for finite nucleus at its 
density is almost equal to the compressibility at similar density of 
finite nucleus. Similarly the binding energy obtained for finite nucleus 
can be equated with the nuclear matter value at the particular density 
of finite nuclei. This is also compared in Table I. Analogous to the 
compressibility, the binding energy of a nucleus can also be expressed 
in terms of a leptodermous expansion \cite{meye}:
\begin{eqnarray}
BE(A,Z)&=&a_v A - a_{s}A^{2/3}-a_{coul}\frac{Z(Z-1)}{A^{1/3}} \nonumber \\
&&-a_a\frac{(N-Z)^2}{A}+....,
\end{eqnarray}
where, $a_v$, $a_s$ and $a_{coul}$ have their usual meaning of volume, 
surface and Coulomb coefficients, respectively. When we switch off the 
Coulomb repulsion for a symmetric finite nucleus, like $^{40}$Ca, the 
binding energy comes out from the volume and surface contributions. For 
example, the total binding energy of $^{40}$Ca is 342.216 MeV (8.5554 
MeV per particle) in an extended Thomas-Fermi calculation. It is 319.388 
MeV in Thomas-Fermi level, i.e., without surface correction (7.9847 MeV 
per particle). The contribution comes from Coulomb repulsion due to the 
20 protons is 83.714 MeV, i.e., 4.1857 MeV per proton. Then the binding 
energy per nucleon only from volume contribution is (8.5554 + 0.5707 + 
4.1857) MeV = 13.3118 MeV, of course the quantal effect is neglected in 
the evaluation. These values are ($7.596812+0.5095155+3.277564)=11.384$ 
and ($7.8813395+0.53530075+3.468285)=11.885$ MeV for $^{40}$P and 
$^{40}$S, respectively. We get different binding for $^{40}$P, $^{40}$S 
and $^{40}$Ca and do not coincides with the nuclear matter binding as 
shown in Table I. This means, the binding arises from the proton-neutron 
orientation is different than neutron-neutron configuration.
That means, one may not get the binding energy of finite nucleus with 
the help of the leptodermous expansion of mass formula. Because, the
singlet-singlet and triplet-triplet interaction is less attractive than 
the singlet-triplet nucleon-nucleon interaction. Hence, the binding 
energy, very much depends on the nucleon-nucleon configuration both 
in finite nuclei and infinite nuclear matter, i.e., it is not only a 
function of mass number $A$, but function of both proton and neutron 
separately even in the contribution of volume energy in the leptodermous 
expansion. Thus, the leptodermous expansion in the power of nuclear mass 
to obtain the binding energy is only an approximation. The failure of 
the mass formulae to predict the properties of nuclei away from the 
$\beta-$stability line based on leptodermos is noticed by Mittig et al. 
and later on confirmed by many authors \cite{satpathy12}. The root of 
disagreement between $BE/A$ and $^{eos}BE/A$ for a definite $\rho$ may 
be the following:

\begin{itemize}
\item There is no direct relation between $BE/A$ and $^{eos}BE/A$ as 
their sources of origin are different.

\item There is no route to go from $BE/A$ to $^{Es}BE/A)$ and vice verse.
\end{itemize}

The similar situation is also aroused for compressibility $K_A$. The $K_A$
values for $^{40}$P, $^{40}$S and $^{40}$Ca are listed in Table I. The 
difference in compressibility between ETF and TF calculations for $^{40}$P, 
$^{40}$S and $^{40}$Ca are 1.987, 2.048 and 1.976 MeV, respectively. This 
is clear that the surface correction for these nuclei is less than $2\%$.
The giant monopole excitation energy $E_M = {\sqrt \frac{C_m}{B_m}}$, where 
$C_m$ is the restoring force and $B_m$ is the mass parameter. From this 
excitation energy, we have evaluated the compressibility modulus using 
equation (2).  On the other hand the compressibility for nuclear matter 
$K(\rho)$ as a function of density $\rho$ is obtained from the formula 
\cite{blaizot80},
\begin{eqnarray}
K(\rho)=9\rho^2\frac{\partial^2 ({\cal E}/{\rho})}{\partial \rho^2},
\end{eqnarray}
with ${\cal E}$ is the nuclear matter energy density \cite{aru04}. 
This may be due to the similar reasons as it is highlighted for 
binding energies about their different origin and unconnected relations 
of $K_A$ and $K(\rho)$.
\begin{figure}
\begin{center}
\vspace{0.4cm}
\hspace{-0.3cm}
\includegraphics[scale=0.33]{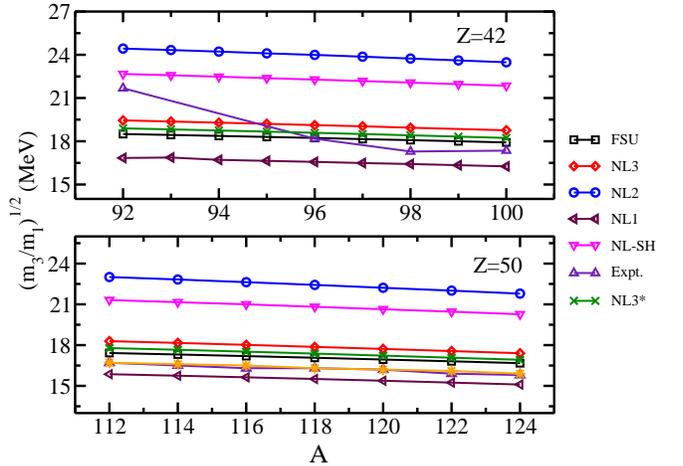}
\caption{\label{fig:epsart} The giant monopole excitation energy obtained 
from various relativistic parameter sets are compared with experimental 
data for Mo and Sn isotopic series. The upper part is for Mo series and 
the lower part is for Sn isotopic series.}
\end{center}
\end{figure}
Now come to the description of giant monopole resonances and their link 
with the nuclear matter compressibility. Recently, hot discussions are 
going on to settle the issue of compressibility as well as its relation 
on giant monopole resonances (GMR). As it is stated earlier, the 
compressibility is not a measurable quantity, but is estimated using 
equation (2), where the GMR energy is an input. The monopole excitation 
energy is measured experimentally, which is a dynamical quantity. However, 
the compressibility $K_A$ or $K_{\infty}$ is a static quantity. Hence, it 
is always a challenge to estimate a static quantity from a dynamical source 
and we apprehense to get an accurate compressibility from the monopole 
excitation energy.

To verify this, we have plotted the sum rule weight factor 
$\frac{m_3}{m_1}$, which is the monopole excitation of the giant 
resonances in Figure 1 for Mo and Sn isotopes. Different force 
parameters having a wide range of nuclear matter compressibility 
from $K_{\infty}=$ 210 MeV to 400 MeV are deployed in the calculations. 
For example, $K_{\infty}\approx$210 MeV for NL1 \cite{rufa86} and 
$K_{\infty}\approx$400 MeV for NL2 force \cite{lee87}. The compressibility, 
represented in the parenthesis, lies in between these two extremes for 
all other [FSUGold (230 MeV), NL3 (271 MeV) \cite{lala97}, NL-SH (355 MeV) 
\cite{sharma} and NL3*(258 MeV) \cite{lala09}] parameter sets. We have 
also displayed the experimental results for comparisons. Although, the 
calculated results are obtained from a wide range of parameter sets having 
variety of $K_{\infty}$, non of the parametrization could reproduce the 
experimental data for the whole isotopic series. The FSUGold parameter 
set is able to reproduce the data for Pb isotopes (See Table II) at an 
excellent agreement with the experimental values, as the set is designed 
to reproduce the GMR data for Pb and few other nuclei, however it fails 
to reproduce the data for Sn and Mo isotopes (see Fig. 1). For Sn series 
(lower part of the figure), as this set is designed to reproduce the GMR 
for Sn nuclei, however, fails to predict the data of Mo isotopes (upper 
part of the figure). That means, the FSUGold set reproduce the GMR values 
of Pb in an excellent agreement with data, but deviate by $\sim 1$ MeV 
for Sn series and drastically differs for Mo isotopes as shown in Fig. 1 
and Table II.

In summary, in the present letter, we have shown the link of microscopic
calculations with their classical counter parts including the leptodermous
expansion for various physical observables. We have shown that the binding 
energy and compressibility modulus deduced from infinite nuclear matter can 
be approximated to the finite nucleus observables to some extend. There is 
a large discontinuity in the bridge between these two quantities and not 
possible to reach from one end to the other. That means, one can not get 
the finite nucleus binding energy and compressibility knowing the physical 
quantities in nuclear matter condition. Thus, the leptodermous expansion 
for compressibility and mass model are merely formulae and have no physical 
merit for prediction to unknown territory.

\end{document}